\documentclass[twocolumn]{aa}
\usepackage{color}
\usepackage{epsfig}
\def\simlt{\ \raise -2.truept\hbox{\rlap{\hbox{$\sim$}}\raise5.truept   %
\hbox{$<$}\ }}
\def\simgt{\ \raise -2.truept\hbox{\rlap{\hbox{$\sim$}}\raise5.truept   %
\hbox{$>$}\ }}                                                          %

\def\be{\begin{equation}}
\def\ee{\end{equation}}
\def\newline{\hfil\break}

\def\la{\mathrel{\hbox{\rlap{\hbox{\lower4pt\hbox{$\sim$}}}\hbox{$<$}}}}
\def\ga{\mathrel{\hbox{\rlap{\hbox{\lower4pt\hbox{$\sim$}}}\hbox{$>$}}}}

\def\szth{SZ$_{th}$~}
\def\szkin{SZ$_{kin}$~}

\def\ksm{{km~s$^{-1}$~Mpc$^{-1}$}}
\begin{document}

\title{Studying the leptonic structure of galaxy cluster atmospheres
from the spectral properties of the SZ effect}

   \author{S. Colafrancesco \inst{1,2,3}, D. Prokhorov \inst{4,5} and V. Dogiel \inst{6} }

   \offprints{S. Colafrancesco}

\institute{ ASI Science Data Center, ASDC c/o ESRIN,
            via G. Galilei 00044 Frascati, Italy.
\and
            ASI, Viale Liegi 26, Roma, Italy
\and
            INAF - Osservatorio Astronomico di Roma,
            via Frascati 33, I-00040 Monteporzio, Italy.\\
            Email: cola@mporzio.astro.it
\and
             Moscow Institute of Physics and Technology,
             Institutskii lane, 141700 Moscow Region, Dolgoprudnii,
             Russia.
\and
             Institut d'Astrophysique de Paris, CNRS, UMR 7095,
             Universit\'{e} Pierre et Marie Curie, 98bis Bd Arago, F-75014
             Paris, France.
\and
              Lebedev Physical Institute, 117924 Moscow, Russia.
             }

\date{Received: 5 May 2008/ Accepted: 23 October 2008}

\authorrunning {S. Colafrancesco et al.}

\titlerunning {Spectra of the SZE in galaxy clusters}

\abstract{We study the energetics of galaxy cluster atmospheres by analyzing
the SZ effect spectra around the crossover frequency. We calculated
analytically the expressions of both the crossover frequency and the spectral
slope of the SZE around the crossover frequency in various cases: a thermal
electron population; a power-law, non-thermal, electron population; and a
population of electrons experiencing a stochastic acceleration.
We find that the value of the crossover frequency $X_0$ of the SZE depends
significantly on the cluster peculiar velocity $V_r$ which determines the
amplitude of the kinematic SZE), while the value of the slope of the SZE does
not depend on the kinematic SZE spectrum in the optimal frequency range around
the crossover frequency of the thermal SZE, i.e. in the frequency range $x =
3.5 - 4.5$.
Therefore, while the amplitude of the \szkin produces a systematic bias in the
position of the crossover frequency $X_0$, it does not affect significantly the
spectral slope of the SZE.
We therefore propose using measurements of the spectral slope of the SZE to
obtain unbiased information about the specific properties of various electron
distributions in galaxy clusters as well as in other cosmic structures in which
a SZE can be produced.

}

 \maketitle

\section{Introduction}
 \label{sect.intro}

The SZ effect (hereafter SZE) is a promising tool for studying the complex
physics of cluster atmospheres because it is sensitive to the specific features
of the spectra of the various electron populations responsible for the CMB
photon Comptonization (see e.g. \cite{Colafrancesco2007} for a review).
A SZE can be produced, in fact, by different types of leptonic plasmas in
astrophysical environments, such as: thermal (hot and/or warm) electrons in the
atmospheres of galaxies and galaxy clusters
(\cite{Birkinshaw1999,Itohetal1998,Colafrancescoetal2003}); non-thermal (and
relativistic) electrons in clusters and in the cavities produced by AGN radio
lobes (\cite{EnsslinKaiser2000,Colafrancescoetal2003,Colafrancesco2005});
secondary electrons produced by Dark Matter annihilation in cosmic structures
(\cite{Colafrancesco2004}), in addition to the kinematic SZE related to the
bulk motion of these plasmas.\\
For the majority of galaxy clusters the most prominent form of SZE is the
thermal one, which provides a CMB intensity change of
 \be
 \Delta I_{th} = {2(kT_{\rm 0})^3 \over (hc)^2} y_{\rm th} g(x)
  \label{eq.deltai_th}
 \ee
as produced by Inverse Compton Scattering (ICS) of CMB photons off the thermal
electron population residing in the cluster atmosphere with a Comptonization
parameter $y_{th} = (\sigma_T/m_e c^2)\int d \ell n_e k_B T_e$ given in terms
of the electron plasma number density $n_e$ and temperature $T_e$ (see e.g.
\cite{Birkinshaw1999} for a review and \cite{Colafrancescoetal2003} for a
general derivation) and in terms of fundamental constants, namely the Thomson
cross-section $\sigma_T$, the electron mass $m_e$, the speed of light $c$, the
Boltzmann constant $k_B$ and the Planck constant $h$.
The function $g(x)$ depends on the adimensional frequency $x \equiv h \nu /k_B
T_{0}$, where $T_0 = 2.726$ K is the CMB temperature, containing all the
spectral information about the SZE (see e.g. \cite{Colafrancescoetal2003} for
details).\\
There are three basic spectral features that characterize the thermal SZE
signal:\\
i) a minimum in its intensity located at a frequency
 \begin{eqnarray}
 x_{th,min} & \approx &2.265(1-0.0927 \theta_e + 2.38 \theta_e^2) \nonumber \\
            & & + \tau (-0.00674 + 0.466 \theta_e) \; ,
 \label{eq.xmin}
 \end{eqnarray}
where $\theta_e \equiv k T_e / m_ec^2$, and $\tau= \sigma_T \int d \ell n_e$ is
the optical depth of the electron plasma, whose exact position depends weakly
on the electron spectrum (i.e. on the electron temperature $T_e$ and number
density $n_e$) and equals $\sim 2.26$;\\
ii) a crossover frequency, $X_0$, whose value depends on the electron
pressure/energy density and the electron optical depth
 \be
 X_{th,0} \approx a(T_e) + \tau b(T_e)
 \label{eq.x0}
 \ee
with $a(T_e)= 3.830(1+1.162 \theta_e -0.8144 \theta_e^2)$ and $b(T_e)= 3.021
\theta_e-8.672\theta_e^2$, and is found at a frequency $> 3.83$ for increasing
values of $T_e$ (the value $X_{th,0}=3.83$ is found in the non-relativistic
limit, or in the limit $T_e \to 0$);\\
iii) a maximum of its intensity whose frequency location is
 \begin{eqnarray}
 x_{th,max} & \approx & 6.511(1+2.41\theta_e -4.96\theta_e^2) \nonumber \\
            & & +\tau(0.0161+8.16\theta_e -35.9\theta_e^2)
 \label{eq.xmax}
 \end{eqnarray}
and depends sensitively on the nature of the electron population and on its
energy (momentum) spectrum (see \cite{Dolgovetal2001} for the case of electrons
with a thermal spectrum; see also \cite{Colafrancescoetal2003} and
\cite{Colafrancesco2004,Colafrancesco2005,Colafrancesco2007} for the case of
electron populations with different spectra).

Since the frequency location of the zero in every SZE signal contains a crucial
dependence on the pressure (or energy density) of the electron population (see
e.g. Eq. \ref{eq.x0} for the thermal case; see also \cite{Colafrancescoeta2003}
for the discussion of the general cases), it has been widely proposed to use
this property to measure e.g. the cluster temperature (for a thermal electron
population) or the pressure/energy density in the case of a more general
electron population.

However, observations of the SZE close to the crossover frequency are biased
due to the presence of the kinematic SZE
 \be
 \Delta I_{kin} = - {2(kT_{\rm 0})^3 \over (hc)^2} \tau { V_r \over c} h(x)
 \label{eq.deltai_kin}
 \ee
(see \cite{SunyaevZeldovich1980}; see also Eq. \ref{eq.DI_kin} below), which is
associated with the peculiar velocity $V_r$ of the cluster.
The unknown value of $V_r$ therefore limits the ability to measure the cluster
temperature (or more generally its energy density) directly through the
displacement of the crossover frequency $X_0$ of the SZE in the correct
relativistic treatment.
The possible presence of multiple sources of SZE in galaxy clusters (e.g. the
thermal SZE, the SZE due to an additional warm gas component, the non-thermal
SZE due to a relativistic plasma, the DM-induced SZE; see
\cite{Colafrancesco2007} for a recent review) could provide further
uncertainties and biases in the measurement of the crossover frequency $X_0$ of
the dominant thermal SZE, since they add to the thermal SZE with specific
spectral shapes and amplitudes of their signals.\\
This parameter degeneracy might be broken - in principle - by observing the SZE
signal at many frequencies, a procedure that is limited, however, by the
observational setup of the current SZ experiments operating in a limited number
(usually $\sim$ 3-4) of frequency bands of relatively wide frequency width
($\sim 15 - 30 \%$ of the band central frequency) requiring precise
inter-calibration.

In this paper we explore the possibility of extracting unbiased information
about the nature of the electron population that produces the SZE from the
spectral shape of the relative SZE spectrum.

One of the frequency region where large variation of the SZE spectra are found
is around the crossover frequency of the thermal SZE, at frequency $\sim 220$
GHz (see Fig.1 in \cite{Colafrancesco2007}). Therefore, this is a favorable
region for performing a spectral analysis of various SZE components with the
aim to separate the contributions of various electron populations.\\
To provide quantitative estimates of our proposal that could be tested by SZE
experiments with spectroscopic capabilities, we will present an analytical
derivation of the spectral slope of various sources of SZE around the crossover
frequency of the thermal SZE, i.e. in the range $x= 3.5 - 4.5$ (corresponding
to the frequency range $\sim 200 - 255$ GHz).
We show that, in this frequency region, the SZE spectral slopes are mostly
independent of the cluster peculiar velocity, at variance with the value of the
crossover frequency $X_0$, that is inevitably biased by the unknown value of
the cluster peculiar velocity $V_r$.
For the sake of illustration, we first discuss in Sect. 2 the cases of the
non-relativistic and relativistic formulations of the thermal SZ effect. In
Sect. 3 we discuss the case of a completely non-thermal, relativistic SZE for a
power-law electron spectrum. In Sect. 4 we discuss the physical case of the SZE
produced by a Maxwellian electron distribution experiencing a stochastic
acceleration process. We discuss our results and draw our conclusions in Sect.
5.\\
Throughout the paper, we use a flat, vacuum-dominated, CDM cosmology with
H$_0$=70 \ksm, $\Omega_{M}$=0.3, $\Omega_{\Lambda}$=0.7.


\section{The spectrum of the thermal SZE around the crossover frequency}
 \label{sect.szeslope}

The spectra of the thermal SZE in the relativistic and non-relativistic limits
differ quite significantly around the crossover frequency for increasing values
of thermal plasma temperature (see e.g. \cite{Birkinshaw1999} and references
therein). This fact has been widely indicated as a possible means of extracting
physical information, such as e.g. the electronic plasma temperature $T_e$ from
a measurement of the crossover frequency $X_0(T_e)$ of the SZE spectrum, if the
cluster peculiar velocity $V_r$ is known to a sufficiently high accuracy.
However, cluster peculiar velocity information is often uncertain or even
missing, which means that this technique can be affected by strong systematic
biases.\\
In the following, we show that measuring the spectral slope of the SZE spectrum
around the crossover frequency provides a means of extracting information on
the electron distribution that is independent of the cluster peculiar velocity.

We compare the spectral slopes of the thermal SZE, \szth, and the kinematic
SZE, \szkin, calculated for an isothermal cluster in the frequency range
200-255 GHz, corresponding to the range $x\approx 3.5 - 4.5$ of the
adimensional frequency $x \equiv h \nu/k T_{0}$.

%
We begin, for the sake of clarity, our discussion from the non-relativistic
case where the Kompaneets (1957) equation can be solved analytically and leads,
following Eq. (\ref{eq.deltai_th}), to an analytic expression for the CMB
intensity change in the non-relativistic regime
 \be
 \Delta I_{th,nr} =
{2(kT_0)^3 \over (hc)^2} \tau {kT_e \over m_ec^2} g_{nr}(x)
 \label{eq.DI_nr}
 \ee
with
 \be
 g_{nr}(x) = {x^4 e^{x} \over (e^{x} - 1)^2} \bigg( x \cdot {e^{x} + 1 \over e^{x} - 1}
 -4\bigg)\, .
 \label{eq.g_nr}
 \ee
The subscript $'nr'$ denotes the fact that the previous expression was obtained
in the non-relativistic limit.

The kinematic SZE intensity change (Sunyaev \& Zeldovich 1980), due to the bulk
motion of the cluster gas with peculiar velocity $V_r$ along the los, writes as
 \be
 \Delta I_k = - 2{(kT_0)^3 \over (hc)^2} \tau { V_r \over c} h(x)
 \label{eq.DI_kin}
 \ee
with
 \be
 h_{nr}(x) = {x^4 e^{x} \over (e^{x} - 1)^2} \, .
 \label{eq.h_nr}
 \ee
In the following, we use the quantity $i(x)$ to represent the adimensional CMB
intensity spectral distortion
 \be
 i_{th,k}(x) = {\Delta I_{th,k}  \over  \bigg[ {2(kT_0)^3 \over (hc)^2} \tau \bigg]}
 \, ,
 \label{eq.i(x)}
 \ee
which contains relevant spectral information on the thermal and kinematic SZE.

The total SZE intensity in the case of a cluster with a single thermal electron
population moving along the line of sight with a peculiar velocity $V_r$ is due
to the combination of \szth and \szkin and is given by
 \be
 i(x) = {kT_e \over m_ec^2} \bigg[ g(x) - {V_r \over c} {m_ec^2 \over kT_e} h(x)
 \bigg] \, .
 \label{eq.i_th+k}
 \ee
The series expansions of $g(x)$ and $h(x)$ around the adimensional frequency
$x=4$ (see Appendix A) provides the values of the spectral slope $S$ and of the
crossover frequency $X_0$.\\
The slope of the SZE spectrum around the crossover frequency $X_0$ is defined
by the quantity
 \be
 S = {i(x)-i(X_0) \over x - X_0} = {i(x) \over x - X_0}
 \ee
\begin{figure}
 \begin{center}
 \hspace{-0.5cm}
 \epsfig{file=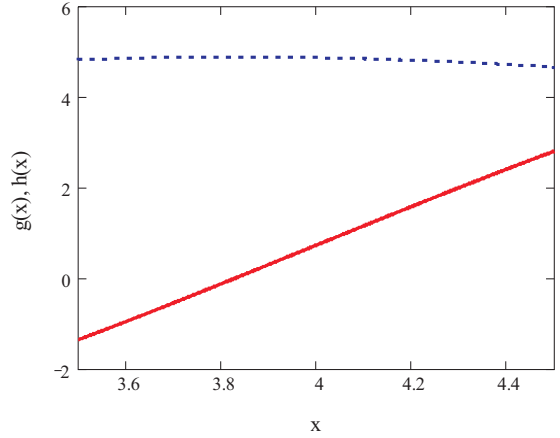,height=6.cm,angle=0}
 \caption{The functions $g_{nr}(x)$ (solid) and $h_{nr}(x)$ (dashed) are shown
 in the frequency range $x=3.5 - 4.5$.}
 \end{center}
 \label{sz_spectra}
\end{figure}
In Fig. 1,  we compare the functions $g(x)$ and $h(x)$ in the frequency range
of our interest. Since the \szkin spectral shape $h(x)$ is almost flat in the
frequency range $x \approx 3.5 - 4.5$, the slope $S$ of the thermal SZE can be
used to extract unbiased information about the nature of the electron
population, i.e., in this case, its temperature $T_e$.

%
In the non-relativistic limit, we find (see the Appendix for a derivation) that
the slope of the total SZE (thermal plus kinematic) is given by
 \be
 S_{nr} = {kT_e \over m_ec^2} {g(X_0 + \delta x) \over \delta x} \approx  4.25
 {kT_e \over m_ec^2} = 0.08 {kT_e \over 10 keV} \,
 \label{eq.slope_nr}
 \ee
and does not depend on the cluster peculiar velocity but only on the electron
temperature $T_e$.\\
The crossover frequency of the total SZE
 \be
  X_{0,total,nr} = 3.83 + 0.193 \bigg({ V_r \over 10^3 km s^{-1}} \cdot {10 keV \over
  kT_e}\bigg) \, .
   \label{eq.xo_th_nr}
 \ee
obtained from the condition $i(X_0)=0$ (see Appendix A.1) depends, instead, on
both $kT_e$ and $V_r$, so that a measure of the crossover frequency for the
thermal SZE cannot provide an unambiguous estimate of the cluster temperature
$T_e$.
Specifically, for values  $kT_e = 10$ keV and $V_r = \pm 1000$ km s$^{-1}$, the
values of $X_{0,total,nr}$ is found to be in the range $3.64 - 4.02$, while for
$V_r=0$ one recovers the non-relativistic value $X_{o,total,nr}=3.83$, which,
obviously, does not depends on the cluster temperature.

%
In the relativistic treatment, the CMB spectral distortion due to the thermal SZE is
given by the expression
 \be
\Delta I_{th} = {2(kT_{0})^3 \over (hc)^2} \cdot  \tau
 \int ds P_1(s) \bigg( {x^3 e^{-3s} \over \exp(x e^{-s})- 1} -
 {x^3 \over e^{x} - 1} \bigg)
 \label{eq.dith}
 \ee
where
 \be
 P_1(s) = \int d \beta f_e(\beta) P(s, \beta)
 \label{eq.P1}
 \ee
with $f_e(\beta)$ being the velocity spectrum of the electron population and
$P(s, \beta) ds$ being the probability that a single scattering of a CMB photon
off an electron with speed $\beta c$ causes a logarithmic frequency shift $s
\equiv ln(\nu ' / \nu$). Here, we use the first order approximation in $\tau$
of the photon redistribution function $P(s)$ for the sake of simplicity.
Analogous conclusions hold, nonetheless, for the more general derivation (see
e.g. \cite{Colafrancescoetal2003}).

\begin{figure}[ht]
\centering
   \epsfig{file=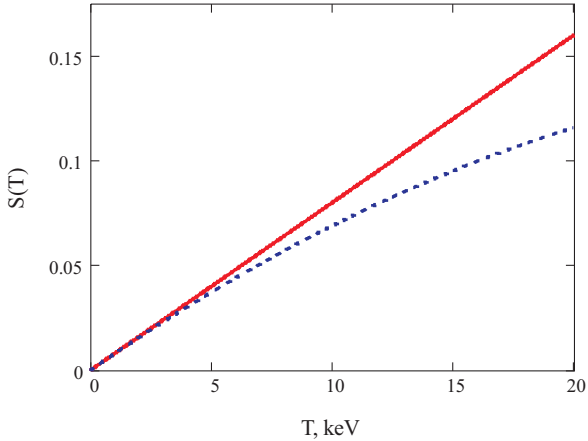,height=6.cm,angle=0}
      \caption{The spectral slopes of the non-relativistic SZE (solid curve) and of the
      relativistically correct SZE spectrum (dashed curve) calculated in the range $x=3.5-4.5$
      for a thermal electron population are shown as a function of the electronic temperature $T_e$
      given in keV.
      }
  \label{szth_slope}
  \end{figure}

The series expansion of $i(x)$ around the crossover frequency $X_0$ was
calculated in Eq. (A.11) of the Appendix.
Using the expression for the crossover frequency
$X_0(T_e)=3.830(1+1.162\theta_e - 0.8144 \theta_e^2)$ given by
\cite{Dolgovetal2001}, one can calculate the slope of the thermal SZE in the
relativistic treatment as a function of $kT_e$.
A good approximation of the slope $S$ in the relativistic treatment, and in the
temperature range $kT_e= 1 - 20$ keV,  is given by
 \begin{eqnarray}
 S_{rel}  & = & 4.25 {kT_e \over m_ec^2} - 30 \bigg({kT_e \over m_ec^2}\bigg)^2 \nonumber
 \\
 & = &
 0.08 {kT_e \over 10 keV} \bigg( 1 - 0.138 {kT_e \over 10 keV} \bigg) \,
 \label{eq.slope_rel}
 \end{eqnarray}
and, again, depends only on the cluster temperature $T_e$.
The slopes of the non-relativistic and relativistically correct thermal SZE
spectrum are shown in Fig. \ref{szth_slope} for comparison.
The value of the crossover frequency of the total SZE in the relativistic
treatment is given by
 \begin{eqnarray}
 X_{0,total,rel} & = & X_{0}(T_e)+\frac{V_{r}}{c}\frac{h[X_{0}(T_e)]}{S_{rel}} \nonumber
 \\
 & \approx &
 X_{0}(T_e) + \frac{0.193}{1-0.138 \cdot {kT_{e} \over 10keV}} \nonumber \\
 & \times &
 \bigg({ V_r \over 10^3 km s^{-1}} \cdot {10 keV \over kT_e}\bigg) \;
 \label{eq.xo_th_rel}
 \end{eqnarray}
(see the Appendix for a derivation) where $X_0(T_e)=3.83 (1 + 1.162 \cdot
\theta_e - 0.814 \cdot \theta_e^2)$.
These results show that, while the value of the crossover frequency
$X_{0,total,rel}$ depends on both $kT_e$ and on $V_r$ (see Eqs.
\ref{eq.xo_th_rel} and \ref{xo_th_rel}), the value of $S$ does not depend on
$V_r$ in the frequency range around the crossover frequency $X_0$ (see Appendix
A for details).
This is because the slope of the kinetic SZE is approximately zero in this
frequency range. Therefore, a measure of $X_0$ for the thermal SZE cannot
provide an unambiguous estimate of the cluster temperature. In contrast, a
measure of the slope of the thermal SZE can provide unbiased constraints on the
value of the electron temperature and, hence, the nature of the electron
plasma.\\
In the following sections, we demonstrate that this is a general result for the
SZE that can be applied to the case of other electron plasmas residing in
cluster atmospheres.

\section{The spectrum of the non-thermal SZE}
 \label{sect.powerlaw}

We consider a non-thermal electron population with a power-law spectrum of the
form
 \begin{eqnarray}
 f_e(p) =& {\alpha - 1 \over p_1^{1 - \alpha} - p_2^{1- \alpha}} \cdot  p^{-\alpha}
            & p_1 < p < p_2 \nonumber \\
 f_e(p) =&  0 & {\rm elsewhere}
 \label{eq.fp}
 \end{eqnarray}
where $f_e(p)$ is the momentum distribution, and the momentum $p = \beta
\gamma$ is normalized to unity. This electron spectrum is expressed as a
function of $\beta$ by
 \be
 f_e(\beta) = {\alpha - 1 \over p_1^{1-\alpha} - p_2^{1-\alpha}} \cdot
 \bigg( {\beta \over \sqrt{1 - \beta^2}} \bigg)^{-\alpha} \cdot \bigg({1 \over 1 - \beta^2}
 \bigg)^{3/2} \, .
 \label{eq.fbeta}
 \ee
\begin{figure}
 \begin{center}
 \epsfig{file=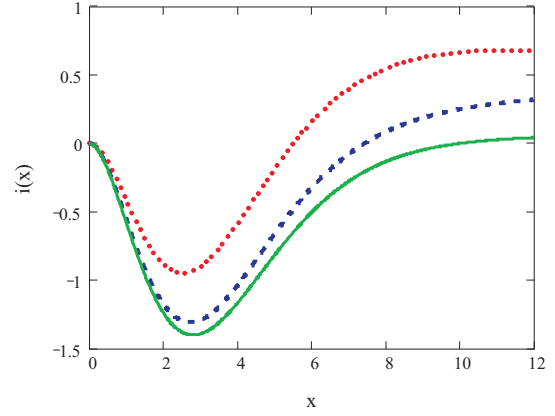,height=6.cm,angle=0}
 \caption{The spectral shape of the normalized CMB intensity change $i(x)$
 evaluated for a power-law electron spectrum with a minimum momentum $p_1=1$ (dotted line),
 3 (dashed) and 10 (solid).}
 \end{center}
 \label{fig.idx}
\end{figure}
The spectral shape of the quantity $i(x)$ evaluated for the previous power-law
electron spectrum is shown in Fig. \ref{fig.idx} for different values of the
minimum momentum $p_1$ of the electron spectrum with a value of the momentum
$p_{2}=1000$.\\
The approximate value of the crossover frequency of the non-thermal SZE for a
monoenergetic electron spectrum in the limit $\gamma\gg1$ has been derived in
the Appendix, and is given by
 \be
 X_{0,non-th} \approx 2 ln\bigg( {4 \over 3} \gamma^2 \bigg) \, .
 \label{eq.xo_pl}
 \ee
The dependence of $X_{0,non-th}$ on the lower momentum $p_1$ of the power-law
electron spectrum is shown in Fig. \ref{fig.x0_pl} (this figure show the values
of $X_{0,non-th}$ without the effect of the peculiar velocity).
\begin{figure}
\centering
 \epsfig{file=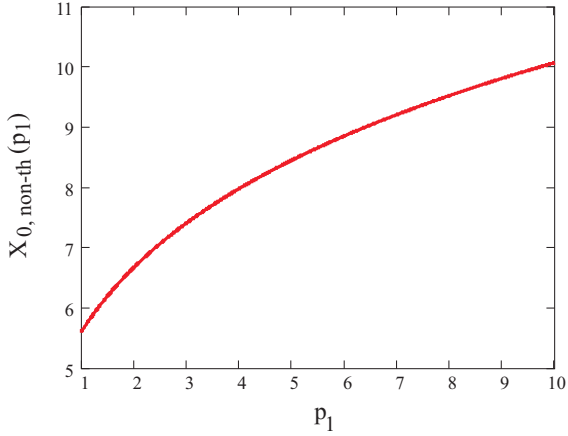,height=6.cm,angle=0}
 \caption{The dependence of $X_{0,non-th}$ on the lower cutoff momentum $p_1$.
 We do not consider here the effect of a possible kinematic SZ effect.}
 \label{fig.x0_pl}
\end{figure}

The quantity $X_{0,non-th}$ is assumed to have values larger than $x \simgt 5$
for all values $p_1 \geq 1$ (see Fig. \ref{fig.x0_pl}). This means that for a
system containing only a non-thermal electron population, the value of
$X_{0,non-th}$ (and hence of the energetics/pressure of the relative electron
population) cannot be determined without a precise knowledge of its peculiar
velocity $V_r$.

The value of the slope of the non-thermal SZE is also affected by the presence
of a kinematic SZE in this case. However, both the \szkin and primary CMB
anisotropies have a flat spectral shape in the frequency range $x = 3.5 - 4.5$,
so that their contribution to the slope of the total SZE is negligible in this
frequency range.

\subsection{Combination of thermal and non-thermal SZE}

For the case of realistic clusters with radio halos and/or X-ray cavities (i.e.
those clusters definitely containing a population of relativistic electrons
whose energy spectrum is often represented by a power-law distribution, see
\cite{Colafrancesco2007} for a review), there is usually a cospatial
distribution of thermal and non-thermal plasmas.
We calculate the values of the cross-over frequency $X_0$ and the slope $S$ of
the SZE for a combination of a thermal plus a power-law electron spectra with
the addition of a peculiar velocity of the cluster.

In this case, we found that the value of the crossover frequency $X_0$ depends
on the cluster temperature, the peculiar velocity, and the optical depths of
the thermal and non-thermal electron populations
 \begin{eqnarray}
X_0 & = & 3.83 + 0.193 \bigg( {V_r \over 10^3km/s} {10keV \over
kT_e} \bigg)
\nonumber \\
    & &
    + 0.148 \bigg( {\tau_{rel} / \tau \over 0.01} {10keV \over kT_e} \bigg)
 \label{eq.x0_pl_th}
 \end{eqnarray}
(see Appendix for details).\\
The value of the slope of the total (thermal plus non-thermal plus kinetic) SZE
is given by
 \begin{eqnarray}
 S  & = & 0.084 {kT_e \over 10 keV} \bigg(1 + 7 \cdot 10^{-3} {V_r \over 10^3km/s}
 {10keV \over kT_e} \nonumber \\
       & & + 0.038 {\tau_{rel} / \tau \over 0.01} {10keV \over kT_e} \bigg)
 \label{eq.slope_pl_th}
 \end{eqnarray}
To estimate specific values of $X_0$ and $S$ for a representative Coma-like
cluster that contains both thermal and non-thermal plasmas, we choose
representative values $kT_e = 8.2$ keV, $p_1 = 10$, and $\tau_{rel} / \tau =
0.01$.\\
For the single Maxwellian spectrum, one obtains in this case the reference
values $X_0 = 3.9$ and $S = 0.058$.
The zero of the total SZE evaluated for a combination of thermal and
non-thermal populations is $X_0 = 4.09$, and the increase in the value of $X_0$
is in agreement with Eq. (\ref{eq.x0_pl_th}). The spectral slope of the total
SZE for a combination of the two electron populations is $S=0.061$, and the
mild increase in $S$ in this case is also in good agreement with Eq.
(\ref{eq.slope_pl_th}).

The contribution of the kinematic SZE to the determination of $X_0$ cannot be
ignored.
Using Eq. (\ref{eq.x0_pl_th}), we calculated the contributions to the value of
$X_0$ of the terms depending on $V_r$ and $\tau_{rel}/\tau$ .
The contribution of $V_{r}$ to the value of $X_{0}$ can be neglected in
contrast to the contribution of the non-thermal electron population only for
values $|V_r| < 776.8 {\rm km s^{-1}} ({\tau_{rel}/\tau \over 0.01})$;
therefore, even for values $(\tau_{rel}/\tau) \sim 0.01$, the value of $X_{0}$
is substantially affected by possible values of $V_{r}$ in the range $\sim 800
- 1000$ km/s, as found in the tail of the peculiar velocity distribution (with
rms value of $\approx 300 \pm 80$ km/s) of galaxy clusters (see
\cite{Giovanellietal1998}.\\
In contrast, (from Eq. \ref{eq.slope_pl_th}), the impact of the kinematic SZE
(i.e. of the cluster peculiar velocity) on the slope can be neglected in
comparison with the impact of the non-thermal electron population, for values
$| V_{r} | < 5.4 \cdot 10^{3} km s^{-1} ({\tau_{rel}/\tau \over 0.01})$; this
means that the slope of the total SZE is almost unaffected by realistic values
of the cluster peculiar velocity which are well below 5000 km/s.


\section{A specific test case: the SZE spectrum of accelerated electrons}
 \label{sect.testcase}

To test the power of the method discussed in this paper, we apply our previous
analysis to the specific case of a cluster in which the SZE signal is produced
by an electron population experiencing a stochastic acceleration process from a
background plasma (see \cite{Dogieletal2007} for an extensive discussion of
this scenario).

To describe the distribution function of the accelerated electrons, we use a
kinetic equation modeling the influence of both Coulomb collision and
stochastic acceleration, which is valid at sub-relativistic and relativistic
energies (see e.g. \cite{Dogiel2000,Liangetal2002} and \cite{Dogieletal2007})
 \be
 {\partial f_e  \over \partial t} - {1 \over p^2} {\partial \over \partial p}
 \bigg( A(p) {\partial f_e \over \partial p} + B(p) f_e \bigg) = 0 \; .
 \label{eq.kin_eq}
 \ee
Here
 \be
 B(p) = p^2 \bigg({dp \over dt} \bigg)_i
 \label{eq.b}
 \ee
and
 \be
A(p) = B(p) {\gamma \over (\gamma^2 - 1)^{1/2}} \bigg({kT_e \over m_e
c^2}\bigg)^{1/2} + p^2 D_p(p) \; ,
 \label{eq.a}
 \ee
where $D(p)=\alpha p^q$ is the diffusion coefficient due to the stochastic
acceleration (see e.g. \cite{Dogieletal2007}.
The rate of ionization losses is
 \begin{eqnarray}
\bigg( {dp \over dt} \bigg)_i = {1 \over p} \sqrt{p^2 + {m_ec^2 \over kT_e}}
{\gamma \over
 (\gamma^2 - 1)^{1/2}} \nonumber \\
 \times \bigg[ \ln \bigg({E(p) m_ec^2 (\gamma^2 - 1) \over h^2 \omega^2_p \gamma^2} \bigg) + 0.43
 \bigg] \, ,
 \label{eq.i_losses}
 \end{eqnarray}
where $\gamma = 1 + E(p) / m_ec^2$, and $\omega_{p}$ is the plasma frequency.
Finally, the electron distribution function takes the form (Dogiel et al. 2007)
 \be
f_e = C \cdot exp \bigg(-\int_0^p dv {B(v) \over A(v)} \bigg) G(p) \; ,
 \ee
where
 \be
 C= {exp(-1/\theta_e) \sqrt{\theta_e} \over K_2(1/\theta_e)}
 \ee
and
 \be
 G(p) = 1 -{ \int^p_0 {dw \over A(w)} exp (\int^w_0 dt B(t)/A(t)) \over
 \int^{\infty}_0 {dw \over A(w)} exp (\int^w_0 dt B(t)/A(t)) } \, .
 \ee
The integral
 \be
\int^p_0 {B(v) \over A(v)} dv = \int^p_0 {v dv \over \sqrt{1+\theta_e v^2}
\bigg(1+\alpha_* \theta_e v^{q+3} ({1 \over 1 + \theta_e v^2})^{3/2} \bigg)}
 \ee
depends on the acceleration parameter
 \be
\alpha_* = \alpha \times  \bigg[ \ln\bigg({E(p) m_ec^2 (\gamma^2 - 1) \over h^2
\omega^2_p \gamma^2} \bigg) + 0.43 \bigg] \,.
 \ee
The velocity distribution of the accelerated electrons is given by
 \be
P_e(\beta, \alpha_*) = {\beta^2 \over \theta_e^{3/2}(1 - \beta^2)^{5/2}} \cdot
f \bigg({\beta \over \sqrt{\theta_e(1 - \beta^2)}}, \alpha_* \bigg) \;.
 \ee
The resulting spectrum of the background and accelerated particles is shown in
Fig. 5.
%
\begin{figure}
 \begin{center}
%
 \epsfig{file=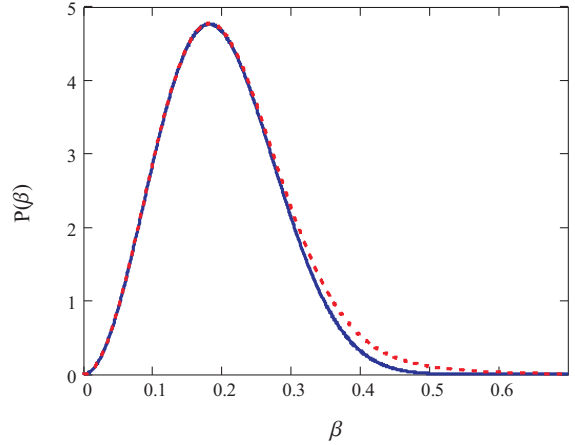,height=6.cm,angle=0}
 \caption{The spectrum of stochastically accelerated electrons with acceleration parameter $\alpha = 0.0008$
 (dashed curve) is compared with the Maxwellian spectrum (solid curve). A background plasma
 temperature of $kT_e=8.2$ keV is assumed here.}
 \end{center}
 \label{fig.e_spectrum}
\end{figure}
The following limiting cases hold:\\
i) if $\alpha_* = 0$, the spectrum is a relativistic Maxwellian spectrum
 \be
P_e(\beta) = {\gamma^5 \beta^2 exp(-\gamma/\theta_e) \over \theta_e
K_2(1/\theta_e)}
 \ee
where $K_2$ is the modified Bessel function of second kind;\\
ii) if $p \ll 1/\sqrt{\theta_e}$, the electron spectrum takes the form of the
non-relativistic spectrum derived by Gurevich (1960)
 \begin{eqnarray}
f_e(p) & = & \sqrt{2 \over \pi} \bigg[ exp \bigg(-\int^p_0 {dw \over 1/w +
\alpha_* \theta_e w^{q+2}} \bigg) \nonumber \\
     & & - exp \bigg(-\int^{\infty}_0 {dw \over 1/w + \alpha_* \theta_e w^{q+2}} \bigg)
     \bigg]\, .
 \end{eqnarray}
The position of the cut-off in the electron spectrum is, in principle, a free
parameter to be fixed at any energy above the range of available X-ray
observations, e.g. at $E > 80$ keV (i.e. the upper value of the energy range of
the Beppo-SAX PDS instrument). In our analysis, we consider specifically a
value of the cut-off energy $E_{cutoff} = m_ec^2(\gamma - 1)$ $= 212$ keV,
corresponding to a value of $p_{cutoff} = 1/\sqrt{\theta_e}$, where $\gamma =
\sqrt{1 + \theta_e p^2}$ that is able to reproduce both the soft and hard X-ray
spectrum of Coma (see Dogiel et al. 2007).
In such a model for the origin of the hard X-ray emission from the Coma
cluster, we derived a value of the acceleration parameter $\alpha = 0.0008$,
given the value of the electron temperature of Coma $k T_e = 8.2$ keV (Dogiel
et al. 2007). These parameters fix the shape of the spectrum of accelerated
electrons and, therefore, the spectral shape of the relative SZE.

\subsection{The dependence of $X_0$ and of $S$ on the acceleration parameter}

We calculate the dependence on the value of $\alpha$ of both $X_0$ and $S$ for
the total SZE of the considered cluster  for the specific model previously
discussed.
For the original Maxwellian spectrum of Coma with $kT_e = 8.2$ keV, we obtain
values $X_0=3.9$ and $S = 0.058$ in the range $x=3.5 - 4.5$.\\
We use the relativistic spectrum of thermal electron population (see Sect. 2)
with values of the acceleration parameter $\alpha$ in the range $\alpha = 0 -
0.0015$, bracketing the value $\alpha = 0.0008$ obtained from the fit to the
hard X-ray spectrum of Coma (Dogiel  et al. 2007).

The approximated expression of the crossover frequency $X_0(\alpha)$ up to
order $\mathcal{O} (\alpha^2)$ is given by
 \be
 X_0(\alpha; k T_e = 8.2 keV) = 3.9 + 15.73 \alpha + 15402 \alpha^2 \; .
 \label{eq.x0_alpha}
 \ee
For the value $\alpha  = 0.0008$, one obtains $X_0 = 3.922$. This value of
$X_0$ would correspond to a thermal electron spectrum (without acceleration) at
an effective temperature of $kT_e=10.5$ keV.\\
We note that in this case the shifts induced on the value of $X_0$ due to both
the electron acceleration process and the presence of a peculiar velocity of
the cluster are intertwined.
\begin{figure}
 \begin{center}
%
 \epsfig{file=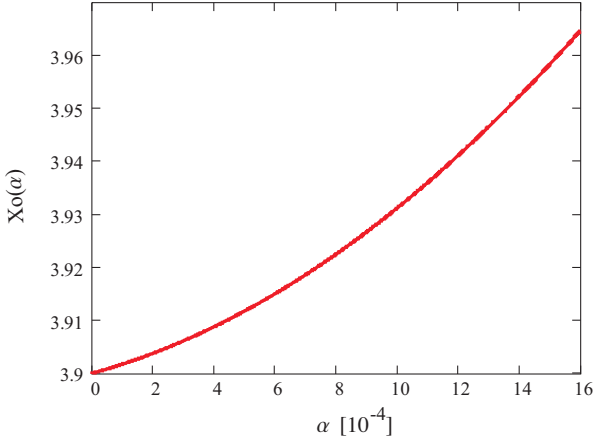,height=6.cm,angle=0}
 \caption{The dependence of the crossover frequency $X_0$ on the acceleration
 parameter $\alpha$ given in units of $10^{-4}$.}
 \end{center}
 \label{sz_x0}
\end{figure}
The expression for the crossover frequency, taking into account the kinematic
SZE, is given by
 \be
 X_{o,total}(\alpha)=
 X_{0}(\alpha)+\frac{V_{r}}{c}\frac{h[X_{0}(\alpha)]}{S(\alpha)} \; .
 \ee
The peculiar velocity of the Coma cluster was estimated to be $V_{r} = -29 \pm
299 km/s$ (Colless et al. 2001), therefore $X_{0,total}$ is in the range
$3.84-4.00$ for a value $\alpha=8\cdot10^{-4}$. Since the values
$X_{0}(\alpha=0)$ and $X_{0}(\alpha=8\cdot10^{-4})$ lie in this range, the
ability to study the nature of a possible population of supra-thermal electrons
directly from the displacement of the crossover frequency is quite limited.

\begin{figure}
\centering
%
 \epsfig{file=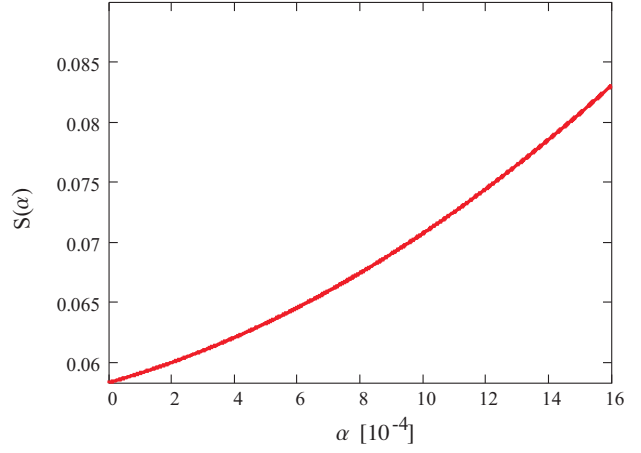,height=6.cm,angle=0}
 \caption{The dependence of the slope $S$ of the SZE in Coma
 on the acceleration parameter $\alpha$ given in units of $10^{-4}$. }
 \label{sz_slope}
\end{figure}

The value of the SZE slope (that, we recall, does not depend on the peculiar
velocity in the frequency range $x = 3.5 - 4.5$), is therefore an important
quantity for deriving useful information about the nature of supra-thermal
electrons.\\
The approximated expression of the quantity $S(\alpha)$ up to $\mathcal{O}
(\alpha^2)$ is given by
 \be
 S(\alpha; kT_e=8.2keV) = 0.0583 + 7.35 \cdot \alpha + 5064 \cdot \alpha^2
 \label{eq.slope_alpha}
 \ee
and its value is $0.0674$ for an electron spectrum with $\alpha = 0.0008$.
The value of $S$ increases by $\approx 16 \%$ with respect to the case of the
Maxwellian spectrum (i.e. with $\alpha = 0$). This value of $S$ would
correspond to a genuine thermal electron spectrum (i.e. without stochastic
acceleration) with an effective temperature of $kT=9.3$ keV. The dependence of
the slope $S$ on the acceleration parameter is shown in Fig. \ref{sz_slope}.

\subsection{Contribution of the kinematic SZE to the slope of the total SZE}

The total SZE intensity in the case of a cluster with an in-situ accelerated
electron population moving along the line of sight with a peculiar velocity
$V_{r}$ is due to the combination of  $SZ_{th}$ and  $SZ_{kin}$
\begin{equation}
i(x)=i_{th}(x)-\frac{V_{r}}{c}h(x) \; .
\end{equation}
The series expansion of the term $SZ_{th}$ around the adimensional frequency
$x=4$ is given by
\begin{equation}
i_{th}=i_{th0}+S \cdot (x-4) + \mathcal{O}(x-4)^2 \; ,
\end{equation}
where $S$ is the value of the slope of the SZE without the contribution of the
kinematic SZ effect.\\
The series expansion of $h(x)$ around $x=4$ is
\begin{equation}
h(x) = 4.865-0.181(x-4)-0.512(x-4)^2 + \mathcal{O}(x-4)^3 \; ,
\end{equation}
which can be written as
\begin{equation}
h(x)=h_{0}+\Delta(x)\cdot(x-4) \; ,
\end{equation}
where $\Delta(x)=-0.181-0.512(x-4)+\mathcal{O}(x-4)^2$.\\
The total SZE intensity is then given by the expression
\begin{equation}
i(x) = i_{th0} - \frac{V_{r}}{c} h_{0} + \left(S_{th} -
\frac{V_{r}}{c}\Delta(x)\right)\cdot(x-4) \; .
\end{equation}
For a narrow frequency range, the spectral slope of the SZE can be derived by
means of SZE measurements at two nearby frequencies, given by
\begin{equation}
S_{obs}=\frac{i(x_{1})-i(x_{2})}{x_{1}-x_{2}}
\end{equation}
For the sake of clarity, we fix the first frequency $x_{1}=4$ and the second frequency
lies in the range $x_{2}=[3.5,4.5]$.\\
In this case, we find that
\begin{equation}
S_{obs} = S_{th} - \frac{V_{r}}{c} \Delta(x_{2}) \; .
\end{equation}
\begin{figure}[ht]
\centering
\includegraphics[angle=0, height=6.cm]{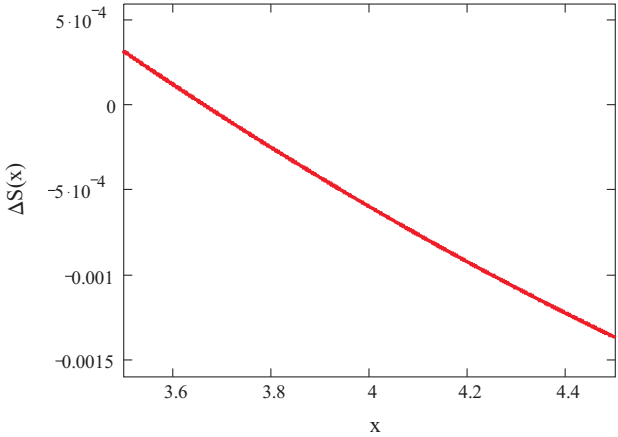}
\caption{The contribution of the kinematic SZE to the slope of the total SZE
(see text for details).}
 \label{kin_con}
\end{figure}
The contribution of the kinematic SZE to the slope of the total SZE is given by
\begin{equation}
S_{kin} = S_{obs} - S_{th} = - \frac{V_{r}}{10^{3} km/s} \Delta S
(x_{2}) \; ,
\end{equation}
where the quantity $\Delta S(x_{2}) =  3.3\cdot 10^{-3} \Delta(x_{2})$ is
plotted in Fig. \ref{kin_con}.\\
It is clear that the contribution of the kinematic SZE to the slope of the
total SZE is far smaller than the slope increase due to the stochastic
acceleration process (see Fig. \ref{sz_slope}). It follows that the
determination of the slope of the SZE is a powerful tool for studying the
presence and the nature of a possible population of accelerated electrons at
supra-thermal energies in the atmospheres of galaxy clusters.


\section{Discussion and conclusions}
 \label{sect.discussion}

We demonstrated in this paper that the value of the crossover frequency $X_0$
of the SZE depends crucially on the cluster peculiar velocity $V_r$ (as well as
on the CMB primary anisotropies), while the value of the slope of the SZE does
not depend on the kinematic SZE spectrum (and on the CMB anisotropy spectrum),
in the optimal frequency range around the crossover frequency of the thermal
SZE, i.e. in the frequency range $x = 3.5 - 4.5$. This is because both the
spectrum of the kinetic SZE and that of the CMB primary anisotropies are almost
flat in this frequency range. Therefore, while the amplitude of the \szkin
produces a systematic bias in the position of the crossover frequency $X_0$, it
does not affect significantly the slope $S$ of the SZE.\\
This fact is more evident when one computes the CMB temperature change produced
by the SZE, i.e. the quantity
 \be
 {\Delta T \over T_{0}} = {{(e^x-1)^2}\over{x^4e^x}} {{\Delta I}\over I_0}\,;
 \label{eq.dt}
 \ee
in fact, the CMB temperature change induced by the kinetic SZE has a flat
spectrum in frequency, while any other CMB temperature change induced by other
SZE have spectra with specific values of slopes that differ from zero.\\
Since the cluster peculiar velocity is unknown in many cases, the precise
position of the crossover frequency $X_0$ of the thermal effect cannot be
well-defined from the SZE observations because the measurements of peculiar
velocity are highly uncertain.

To overcome this systematic bias, we have proposed to use the spectral slope of
the SZE to obtain unbiased information about specific properties of various
electron distributions in galaxy clusters: the gas temperature for a thermal
electron population, the power-law spectrum of a non-thermal electron
population, the combined spectrum of thermal and non-thermal electron
populations, the acceleration parameter of a stochastically accelerated
electron population.
We have evaluated analytically specific values of the slope for various SZE
signals in the frequency range $x = 3.5 - 4.5$.
We found that in this frequency range (around the crossover frequency of the
thermal SZE) the spectral slope of the SZE does not depend on the value of the
cluster peculiar velocity because the spectrum of the kinetic SZE is extremely
flat in this frequency region. We have shown this for different electron
populations residing in clusters.

A doubtless advantage of the method that we have presented here is that it
provides an opportunity to search more reliably for  nonthermal components of
electron spectra in clusters and determine their characteristics. These allow
us to derive independently parameters of acceleration processes in clusters
which are usually unknown and estimated in the framework of different models.
This final circumstance is particularly important for the problem of the origin
of nonthermal emission from clusters in the Extreme UV, in the hard X-ray and
at radio wavelengthes.

Spectroscopic measurements of the SZE spectra will become available in the near
future by means of dedicated space experiments (such as e.g. SAGACE;
Spectroscopic Active Galaxy And Cluster Explorer, see
http://oberon.roma1.infn.it/sagace/) based on microwave spatially resolved
spectroscopic techniques. The advent of these spectroscopic capabilities will
allow us to study in an unbiased way several physical details of the leptonic
structure of the atmospheres of galaxy clusters and other cosmic structures.

\begin{acknowledgements}
The authors acknowledge discussions with P. De Bernardis and S.
Masi. The SAGACE experiment has been selected for Phase-A study by
the Italian Space Agency (ASI) as a small-mission project for the
ASI space program 2008-2012.
DP and VD are partly supported by the RFBR grant 08-02-00170-a, the NSC-RFBR
Joint Research Project ¹ 95WFA0700088 and by the grant of a President of the
Russian Federation "Scientific School of Academician V.L.Ginzburg".
\end{acknowledgements}



\appendix

\section{The shape of the SZE around the crossover frequency}

In this Appendix, we calculate the values of the crossover frequency $X_0$ and
the slope $S$ of the SZE in the frequency range $x = 3.5 - 4.5$. We evaluate
these quantities for various cases of thermal and non-thermal electron
populations with the inclusion of the kinematic SZE.

\subsection{Thermal electron spectrum}


The total SZE intensity, in the case of a cluster with a single thermal
electronic population moving along the line of sight with a peculiar velocity
$V_r$ is due to the combination of \szth and \szkin and is given by
 \be
 i(x) = {kT_e \over m_ec^2} \bigg[ g(x) - {V_r \over c} {m_ec^2 \over kT_e} h(x)
 \bigg] \, .
 \label{eq.i_th+k}
 \ee
In the non-relativistic case, the series expansion of $g(x)$ and of $h(x)$
around $x=4$ (i.e. the central value of the frequency range $x \approx 3.5 -
4.5$) is given by
 \begin{eqnarray}
g(x) & =& 0.726 + 4.28(x - 4) - 0.038(x - 4)^2 - 0.489(x - 4)^3 \nonumber \\
     &  & + \mathcal{O}(x-4)^4
 \end{eqnarray}
and
  \begin{eqnarray}
h(x) & =& 4.865 - 0.181(x - 4) - 0.512(x - 4)^2 + 0.088(x - 4)^3 \nonumber \\
     & & + \mathcal{O}(x-4)^4
 \end{eqnarray}
Thus, the series expansion of $i(x)$ around $x=4$ is given by
 \begin{eqnarray}
 i(x) & = &  {kT_e \over m_ec^2} \bigg[ \bigg( 0.726 - 4.865 {V_r \over c} {m_ec^2 \over
 kT_e}\bigg) \nonumber \\
      & & + \bigg(4.28 + 0.181 {V_r \over c} {m_ec^2 \over kT_e}\bigg) (x - 4) \bigg]
 \nonumber \\
      & = & {kT_e \over m_ec^2} \bigg[ \bigg( 0.726 - 0.827 {V_r \over 10^3 km/s} {10 keV \over
 kT_e}\bigg) \nonumber \\
    & & + \bigg(4.28 + 0.0308 {V_r \over 10^3 km/s} {10 keV \over kT_e}\bigg) (x - 4)
    \bigg] \; ,
 \label{eq.i_total_exp}
 \end{eqnarray}
where we neglect higher order terms in the frequency range $x \approx 3.5 - 4.5$ (i.e.
$\nu \approx  200-255$ GHz).
In the first brackets of Eq. (\ref{eq.i_total_exp}), the term depending on
$V_r$ and $T_e$ cannot be neglected, but, in the second brackets of the same
Eq. (\ref{eq.i_total_exp}) we can neglect the term depending on the combination
$V_r m_ec^2/c k T_e$, since it is always far smaller than the value $4.28$ for
realistic values of $V_r$ and $k T_e$ found for clusters.

The slope, $S \equiv [i(x)-i(X_0)]/(x-X_0) = i(x)/(x-X_0)$ of the
non-relativistic SZE spectrum is
 \be
 S_{nr} = {kT_e \over m_ec^2} {g(X_0 + \delta x) \over \delta x} \approx  4.25
 {kT_e \over m_ec^2} = 0.08 {kT_e \over 10 keV}\, .
 \label{eq.slope_nr}
 \ee
Hence, the slope of the SZE intensity $i(x)$ in this frequency range does not depend on
the cluster peculiar velocity.

The value of $X_0$ of the total (thermal plus kinematic) SZE is defined by the
condition $i(X_{0}) =0$, or
  \be
{kT_e \over m_ec^2} \bigg[ \bigg( 0.726 - 4.865 {V_r \over c} {m_ec^2 \over
kT_e} \bigg) + 4.28 (X_0 - 4)\bigg] = 0 \, .
 \ee
From the previous condition, one obtains
 \be
  X_{o,total,nr} = \bigg(4 - {0.726 \over 4.28} \bigg) + {4.865 \over 4.28} {V_r \over c}
  {m_ec^2 \over kT_e} \; ,
 \ee
which is written, in more practical units, as
 \be
  X_{o,total,nr} = 3.83 + 0.193 \bigg({ V_r \over 10^3 km s^{-1}} \cdot {10 keV \over
  kT_e}\bigg) \, .
 \ee

%
In the relativistic treatment, the CMB spectral distortion due to the thermal SZE is
given by the expression
 \be
\Delta I_{th} = {2(kT_{0})^3 \over h^2c^2} \cdot  \tau
 \int ds P_1(s) \bigg( {x^3 e^{-3s} \over \exp(x e^{-s})- 1} -
 {x^3 \over e^{x} - 1} \bigg)
 \ee
where
 \be
 P_1(s) = \int d \beta f_e(\beta) P(s, \beta)
 \ee
with $f_e(\beta)$ being the velocity spectrum of the electronic population and
$P(s, \beta) ds$ being the probability that a single scattering of a CMB photon
off an electron with speed $\beta c$ causes a logarithmic frequency shift $s
\equiv ln(\nu ' / \nu$). Here, we use the first order approximation in $\tau$
of the photon redistribution function $P(s)$ for the sake of simplicity.
Analogous conclusions hold, nonetheless, for the more general derivation (see
e.g. \cite{Colafrancescoetal2003}.

The series expansion of $i(x)$ around the crossover frequency $X_0$ is written
as
 \begin{eqnarray}
i(x) & = & \int ds P_1(s) \bigg[ {X_0^3 exp(X_0 exp(-s)) exp(-4s) \over (exp(X_0 exp(-s))
-1)^2} \nonumber \\
 & & + {X_0^3 exp(X_0) \over (exp(X_0) - 1)^2} + {3 \over X_0} \bigg({X_0^3 exp(-3s)
 \over exp(X_0 exp(-s)) - 1} \nonumber \\
 & & - {X_0^3 \over exp(X_0) - 1} \bigg) \bigg] (x - X_0) \, .
 \end{eqnarray}
Using the expression $X_0(T)=3.830(1+1.162\theta_e - 0.8144 \theta_e^2)$
\cite{Dolgovetal2001}, one can calculate the slope of the thermal SZE for the
relativistic treatment as a function of $kT_e$.

A good approximation of the slope for the relativistic treatment, and in the
temperature range $kT_e= 1 - 20$ keV,  is given by
 \begin{eqnarray}
 S_{rel}  & = & 4.25 {kT_e \over m_ec^2} - 30 \bigg({kT_e \over m_ec^2}\bigg)^2 \nonumber
 \\
 & = &
 0.08 {kT_e \over 10 keV} \bigg( 1 - 0.138 {kT_e \over 10 keV} \bigg) \, .
 \end{eqnarray}
The value of crossover frequency of the total SZE in the relativistic treatment
$X_{o,total,rel}$ is given by
 \begin{eqnarray}
 X_{o,total,rel} & = & X_{0}(T_e)+\frac{V_{r}}{c}\frac{h(X_{0}(T_e))}{S_{rel}} \nonumber
 \\
 & \approx & X_{0}(T_e) + \frac{0.193}{1-0.138 \cdot {kT_{e} \over 10keV}}
 \nonumber \\
 & \times  & \bigg({ V_r \over 10^3 km s^{-1}} \cdot {10 keV \over kT_e}\bigg) \; .
 \label{xo_th_rel}
 \end{eqnarray}


\subsection{Power-law, non-thermal electron spectrum}
 \label{sect.powerlaw}

For a power-law, electron spectrum given by
 \begin{eqnarray}
 f_e(p) =& {\alpha - 1 \over p_1^{1 - \alpha} - p_2^{1- \alpha}} \cdot  p^{-\alpha}
            & p_1 < p < p_2 \; , \nonumber \\
 f_e(p) =&  0 & {\rm elsewhere}
 \end{eqnarray}
with $p = \beta \gamma$, the electron spectrum in terms of $\beta$ is given by
 \be
 f_e(\beta) = {\alpha - 1 \over p_1^{1-\alpha} - p_2^{1-\alpha}} \cdot
 \bigg( {\beta \over \sqrt{1 - \beta^2}} \bigg)^{-\alpha} \cdot \bigg({1 \over 1 - \beta^2}
 \bigg)^{3/2}
 \ee
If $p_1$ is ultra-relativistic (i.e. $p_1 \gg 1$), the limiting form of $i(x)$
is given by
 \be
 i_{ultra-rel}(x) \to - {x^3 \over e^x -1} \, .
 \ee
One can derive analytically the approximate dependence of $X_0$ on $\gamma$ for
a monoenergetic electron spectrum.
The photon density distribution for a CMB-blackbody spectrum is
 \be
n(x) = {x^2 \over e^x - 1} \; .
 \ee
An electron with Lorentz factor $\gamma$ increases the frequency $x$ of a
scattered photon on average by a factor $x'/x = 4\gamma^2/3$. For relativistic
electrons, the photon is therefore scattered to much higher energies. The
photon spectrum after a scattering is
 \begin{eqnarray}
n'(x') & = & n[x(x')] {dx \over dx'} \nonumber \\
       & = & \bigg({3x' \over 4 \gamma^2} \bigg)^2 {1 \over exp (3x'/ 4 \gamma^2)- 1} \cdot
       {3 \over 4 \gamma^2}
 \end{eqnarray}
The value of $3x' / 4 \gamma^2$ is $\ll 1$ for $\gamma \gg 1$. Since one
expects in this case that $X_0$ is much larger than $1$ because of large
amounts of transferred electron momenta, we can find the value of $X_0$ from
the condition that $exp(x'_0) \gg 1$. In this case, the value of $X_0$ is
defined by the condition
 \be
 n'(x'_0) - n(x'_0) = {9 \over 16 \gamma^4} x'_0 - {(x'_0)^2  \over exp(x'_0)} = 0
 \ee
Since $X_0 \gg 1$, the condition $X_0 \gg ln(X_0)$ holds, and one obtains
 \be
 X_0 \approx 2 ln\bigg({4 \over 3} \gamma^2 \bigg)
 \ee

\subsection{Combination of thermal and non-thermal electron populations}

We evaluate the values of $X_0$ and $S$ for a combination of a thermal plus a
power-law electron spectra with the addition of a peculiar velocity of the
cluster.

One can consider a photon to be removed effectively from the CMB spectrum if
its energy is increased by about one order of magnitude or more, requiring
$\gamma > 3$ (e.g. \cite{EnsslinKaiser2000}. Using the usual Kompaneets
approximation and neglecting those relativistic electrons with $\gamma < 3$,
the produced distorted spectrum is
 \be
i(x) = {kT_e \over m_ec^2} \bigg[ g(x) - {V_r \over c} {m_ec^2 \over kT_e} h(x)
- {\tau_{rel} \over \tau} {m_ec^2 \over kT_e} {x^3 \over e^x - 1} \bigg]
 \ee
The series expansion of $i(x)$ around $x=4$ is then
 \begin{eqnarray}
i(x) & = & {kT_e \over m_ec^2} \bigg[ \bigg( 0.726 - 4.865 {V_r \over c}
{m_ec^2 \over kT_e} - 1.194 {\tau_{rel} \over \tau} {m_ec^2 \over kT_e} \bigg)
\nonumber
\\
  & &+ \bigg( 4.28 + 0.181 {V_r \over c} {m_ec^2 \over kT_e} + 0.321 {\tau_{rel} \over \tau} {
 m_ec^2 \over kT_e} \bigg) (x - 4) \bigg] \nonumber \\
  & &
 \label{combo1}
 \end{eqnarray}
From the condition $i(x)=0$, the value of the crossover frequency $X_0$ of the
total SZE is
 \be
X_0 = 4 - {1 \over 4.28} \times {{0.726 - 4.865 {V_r \over c} {m_ec^2 \over
kT_e} - 1.194 {\tau_{rel} \over \tau} {m_ec^2 \over kT_e} \over {1 + {0.181
\over 4.28} {V_r \over c} {m_ec^2 \over kT_e} + {0.321 \over 4.28} {\tau_{rel}
\over \tau} {m_ec^2 \over kT_e}}}}
 \label{combo2}
 \ee
From Eqs. (\ref{combo1}) and (\ref{combo2}), we then derive the values of
 \begin{eqnarray}
X_0 & = & 4 - {0.726 \over 4.28} + {4.865 \over 4.28} {V_r \over c} {m_ec^2
\over kT_e} + {1.194 \over 4.28} {\tau_{rel} \over \tau} {m_ec^2 \over kT_e}
\nonumber
\\
    & = & 3.83 + 0.193 \bigg( {V_r \over 10^3km/s} {10keV \over kT_e} \bigg)
    \nonumber \\
& + & 0.148 \bigg( {\tau_{rel} / \tau \over 0.01} {10keV \over
kT_e} \bigg)
 \end{eqnarray}
and of the slope of the total SZE
 \begin{eqnarray}
S & = & 4.28 {kT_e \over m_ec^2} \bigg(1 + 7 \cdot 10^{-3} {V_r \over 10^3km/s}
{10keV \over kT_e}  \nonumber \\
  & + & 0.038 {\tau_{rel} / \tau \over 0.01} {10keV \over kT_e}
\bigg) \; .
 \end{eqnarray}


\end{document}